\documentclass[12pt]{article}

\input epsf
\usepackage{amssymb,amsmath,wrapfig}
\usepackage[matrix,arrow,curve]{xy}
\usepackage{cite}
\usepackage{graphicx}

\makeatletter
\@addtoreset{equation}{section}
\makeatother

\def\rr {{\Bbb R}}
\def\cc {{\Bbb C}}
\def\pp {{\Bbb P}}

\def\del {\partial}

\def\ka {K\"ahler}
\def\del {\partial}
\def\tts {{$T\oplus T^*$}}
\def\stt {{$\mathrm{SU(3)}\times\mathrm{SU(3)}$}}
\def\sut {{${\rm SU}(3)$}}

\begin{document}

        \begin{titlepage}

        \begin{center}

        \vskip .3in \noindent

        {\Large \bf{Perturbing gauge/gravity duals by a Romans mass}}

        \bigskip

        Davide Gaiotto$^1$ and Alessandro Tomasiello$^{2,3}$\\

        \bigskip

       {\small $^1$ School of Natural Sciences,
    Institute for Advanced Study, Princeton, NJ 08540, USA\\
    \vspace{.1cm}

        $^2$ Jefferson Physical Laboratory, Harvard University,
        Cambridge, MA 02138, USA
        \vspace{.1cm}

        $^3$ Universit\`a di Milano--Bicocca
        and INFN, sezione di Milano--Bicocca,
        I-20126 Milano, Italy}

        \vskip .5in
        {\bf Abstract }
        \vskip .1in

        \end{center}

        \noindent
We show how to produce algorithmically gravity solutions in massive IIA (as infinitesimal first order perturbations in the Romans mass parameter) dual to assigned conformal field theories. We illustrate the procedure on a family of Chern--Simons--matter conformal field theories that we recently obtained from the ${\cal N}=6$ theory by waiving the condition that the levels sum up to zero.

        \vfill
        \eject


        \end{titlepage}

\section{Introduction} 
\label{sec:intro}

The Romans mass parameter of IIA supergravity \cite{romans-massiveIIA} is understood from a modern perspective \cite{polchinski-dbranes} as the Ramond--Ramond (RR) flux $F_0$. In spite of this, it still retains some aura of mystery. For example, its interpretation in M--theory is still challenging (although see for example \cite{bergshoeff-lozano-ortin,hull-romansmass}). Also, the branes that source it are D8--branes, which have the peculiarity of generating a back--reaction that grows with distance (since there is only one direction transverse to them).

On spaces with boundary conditions with an AdS factor, the AdS/CFT correspondence \cite{maldacena} gives a non--perturbative understanding of string theory. One can then hope to get a non--perturbative understanding of the parameter $F_0$ on such backgrounds. Some non--supersymmetric AdS vacua with $F_0 \neq 0$ were proposed already in \cite{romans-massiveIIA}; supersymmetric ones were found much more recently, starting from \cite{behrndt-cvetic} and more recently in \cite{dewolfe-giryavets-kachru-taylor,t-cp3,koerber-lust-tsimpis}.

It was also anticipated some time ago \cite{schwarz-cs} that vacua with Romans mass would be dual to field theories with a Chern--Simons term. Recently, many Chern--Simons--matter conformal field theories (CFTs) have found their gravity dual in string theory, starting with the ${\cal N}=6$ example on AdS$_4\times \cc\pp^3$ in \cite{aharony-bergman-jafferis-maldacena}. Those gravity duals do not involve the parameter $F_0$. However, it was later shown in \cite{gaiotto-t} that the gauge/gravity duality in \cite{aharony-bergman-jafferis-maldacena} could be deformed by adding $F_0$.

In fact, we found in \cite{gaiotto-t} that several ways of introducing $F_0$ were possible, yielding CFTs with varying amounts of supersymmetry, from ${\cal N}=0$ to ${\cal N}=3$. Two theories, with ${\cal N}=0$ and ${\cal N}=1$, had large flavor symmetries (SO(6) and SO(5) respectively). This helped us find their gravity duals, which were presented already in \cite{gaiotto-t}. The ${\cal N}=2$ and ${\cal N}=3$ theories had smaller flavor symmetry groups, and their gravity dual could not be immediately identified.

In this paper, we partially fill that gap by finding those duals as infinitesimal first--order deformations of the ${\cal N}=6$ solution on AdS$_4\times\cc\pp^3$. To see that the solutions are the right gravity duals, one can at first match the bosonic symmetry group, the amount of supercharges, and the moduli spaces of vacua. One finds, however, that all these matches derive from the match of the abelian superpotential, which actually also guarantees that the solutions are the correct duals, as we will now explain. 

In these backgrounds, even a single D2--brane probe feels an effective superpotential $W$. The D2, then, cannot move freely: it will only preserve supersymmetry along some subspace of $\cc\pp^3$. By AdS/CFT,  $W$ should also be the superpotential for the field theory when the gauge group is abelian. $W$ indeed does not vanish for the ${\cal N}=2$ and ${\cal N}=3$ theories proposed in \cite{gaiotto-t} (unlike for the ${\cal N}=6$ theory of \cite{aharony-bergman-jafferis-maldacena}). In four dimensions, an example of a family of theories whose abelianized superpotential does not vanish is given by the Leigh--Strassler theories \cite{leigh-strassler}; in their gravity dual, D3--brane probes only preserve supersymmetry along some locus. Infinitesimal perturbations of the AdS$_5 \times S^5$ background with these properties have been obtained in \cite{grana-polchinski} at first order and in \cite{aharony-kol-yankielowicz} at second and third order. (For a particular type of Leigh--Strassler theory, the gravity dual can actually be found exactly by solution--generating symmetries \cite{lunin-maldacena}).

After identifying the superpotential felt by a single D2--brane probe with the abelian superpotential of the field theory, it turns out that the first order perturbation\footnote{$F_0$ is quantized in string theory, but it can still be small compared to the other fluxes in the unperturbed solution. In this sense, it makes sense to work in perturbation theory and to postpone consideration of the flux quantization conditions to when one has the full solution.} in $F_0$ of the gravity solution can be found with no extra Ansatz or choice. This is quite general. Suppose one has  a supersymmetric solution with $F_0 =0$, whose CFT dual is known. Suppose one knows that a deformation exists, with a superpotential that does not vanish when abelianized,
and with $F_0 \neq 0$. (The meaning of the latter condition on the field theory side is discussed in \cite{gaiotto-t,fujita-li-ryu-takayanagi}). We observe in this paper that, in such a situation, the conditions for the existence of a supersymmetric deformation of the background, at first order in $F_0$, leave no room to any guesswork. There is a clear procedure that leads to a solution, provided of course one starts with a superpotential which is appropriate for a CFT. This procedure is the AdS$_4$ analogue of \cite{grana-polchinski}, except that there are nontrivial restrictions on the superpotential already at first order. The conditions for AdS$_4$ solutions are more restrictive than the ones for AdS$_5$; for example, the Bianchi identities do not follow from supersymmetry as for AdS$_5$ solutions in IIB \cite{gauntlett-martelli-sparks-waldram-ads5-IIB}.

So, to summarize, we outline a general procedure to deform gauge/gravity duals by an infinitesimal amount of Romans mass $F_0$, and we illustrate it by finding the perturbations of the ${\cal N}=6$ solution on AdS$_4\times \cc\pp^3$ \cite{nilsson-pope,sorokin-tkach-volkov-11-10} dual to the ${\cal N}=2$ and ${\cal N}=3$ theories discussed in \cite{gaiotto-t}. In section \ref{sec:ft} we review those theories. In section \ref{sec:susy} we review the conditions for supersymmetry, and we isolate the function that plays the role of the superpotential for a probe D2--brane. In section \ref{sec:pert} we outline the general procedure for finding infinitesimal $F_0$ perturbations; in section \ref{sec:cp3} we apply it to the ${\cal N}=2$ and ${\cal N}=3$ solutions on AdS$_4\times \cc\pp^3$.


\section{Review of the field theories} 
\label{sec:ft}

Although our procedure is general, to fix ideas we will start by introducing the field theories which will provide its concrete applications in section \ref{sec:cp3}.

In \cite{gaiotto-t}, we introduced several Chern--Simons--matter theories.  The ones of interest for this paper consist of an
${\cal N}=2$ Chern--Simons theory with gauge
group U$(N)\times$U$(N)$, coupled  to
${\cal N}=2$ chiral superfields $A_i, B_i$ and
vector superfields $V_1, V_2$. The action is
\begin{equation}\label{eq:cft2}
\begin{split}
    S=
     \frac {k_1}{4 \pi} S_{{\rm CS},\,{\cal N}=2}(V_1) +&
     \frac {k_2}{4 \pi} S_{{\rm CS},\,{\cal N}=2}(V_2) + \\
    &\int d^4 \theta\,{\rm Tr}(e^{-V_1} A_i^\dagger e^{V_2} A_i +
     e^{-V_1} B_i e^{V_2} B_i^\dagger)
 + \int d^2 \theta\, W_{{\cal N}=2}\ ,
\end{split}
\end{equation}
where
\begin{equation}
    \label{eq:w}
    W_{{\cal N}=2} =
    c_1 {\rm Tr}(B_i A_i)^2 + c_2{\rm Tr}(A_i B_i)^2\ .
\end{equation}
There is a renormalization group flow in the space of the
coefficients $c_1$, $c_2$.
If $k_1 + k_2=0$, there is a fixed point at $c_1=\frac{2 \pi}{k_1}$,
$c_2=\frac{2\pi}{k_2}$: it is the ${\cal N}=6$ of \cite{aharony-bergman-jafferis-maldacena}. If $k_1 + k_2 $ is small but $\ll k_i$, it was argued in \cite{gaiotto-t} that a fixed point will still exist for some value of the $c_i$, although for a different value of the coefficients $c_i$.

In fact, we argued that there is a fixed {\it line}, that passes through two points with enhanced symmetries. In general, the theory (\ref{eq:cft2}) has ${\cal N}=2$ supersymmetry, and
an SU(2) of flavor symmetry (as well as the R--symmetry SO(2)$_{\rm R}$). For
\begin{equation}\label{eq:cn3}
    c_1=\frac{2\pi}{k_1} \ ,\qquad c_2=\frac{2\pi}{k_2}\ ,
\end{equation}
supersymmetry is enhanced to ${\cal N}=3$, and hence we have SU(2)$\times$SO(3)$_{\rm R}$ of R--symmetry. It was argued in \cite{gaiotto-t} that the line of fixed point intersects this locus. Also, for
\begin{equation}\label{eq:n2enh}
    c_1=-c_2
\end{equation}
supersymmetry remains ${\cal N}=2$, but the flavor symmetry gets enhanced to SU(2)$\times$SU(2) ($\times$SO(2)$_R$). The line of fixed points should intersect this locus as well.

General arguments predict \cite{gaiotto-t,fujita-li-ryu-takayanagi}
that the gravity dual of a Chern--Simons--matter theory should have Romans mass
\begin{equation}\label{eq:F0k}
    F_0 = k_1 + k_2 \ .
\end{equation}
 In this paper, we will confirm this picture by finding
the gravity dual to these theories, as an infinitesimal perturbation in $F_0$ of the ${\cal N}=6$ solution on AdS$_4\times\cc\pp^3$. In finding these duals, we have been guided by comparing the superpotential in (\ref{eq:w})
with the superpotential of D2 probes, as we now explain.

Even in the abelian case, the superpotential (\ref{eq:w}) is non--vanishing. In the gravity dual, it should be reproduced by the superpotential felt by a single D2 probe extended along the three--dimensional Minkowski and at fixed radius (in Poincar\'e coordinates). Usually, a single brane probe which is point--like in the internal space and in the radial direction does not feel any superpotential, and the moduli space of its world--volume theory is unrestricted. For example,
for AdS$_5\times$ SE$_5$, where SE$_5$ is a Sasaki--Einstein five--manifold, the moduli space of a D3 probe is the cone over SE$_5$, namely a conical Calabi--Yau, which has real dimension 6. Likewise, for AdS$_4\times$ a tri--Sasaki--Einstein, Sasaki--Einstein, or weakly G$_2$ seven--manifold, the moduli space of an M2 probe is the entire cone over those manifolds -- a conical space with special holonomy and of dimension 8.

Going back to AdS$_5$, an example of background in which a D3 brane probe is {\it not} able to move freely is the Lunin--Maldacena background \cite{lunin-maldacena}, dual to one of the Leigh--Strassler ${\cal N}=1$ gauge theories. In that case, the moduli space of the D3 consists of three copies of $\cc$ intersecting at the origin, which reproduces the fact that, in the field theory, there is a superpotential even at the abelian level. In fact, it is not difficult to show that a D3 brane probe can reproduce this abelian superpotential.

In our three--dimensional field theories we expect a similar phenomenon as in the Leigh--Strassler theories: the abelian version of the superpotential (\ref{eq:w}) will be reproduced in the gravity dual by a D2 domain wall. This fact will help us find the gravity duals: in section \ref{sub:supot} we will derive a general expression for the D2 superpotential in terms of certain differential forms that characterize the gravity solution, which we will now introduce.


\section{Supersymmetry equations} 
\label{sec:susy}
We will review here the conditions for unbroken supersymmetry in the most general setting, using the language of generalized complex geometry.

\subsection{The equations in general} 
\label{sub:susy}

Let us consider a spacetime of the warped--product form
AdS$_4 \times M_6$, which means that the metric is of the form
$ds^2_{10}= e^{2A}ds^2_{{\rm AdS}_4} + ds^2_6$. Then this spacetime is supersymmetric in type IIA\footnote{The conditions for type IIB, that we do not need here, are obtained by $\phi_+\leftrightarrow \phi_-$.} if and only if \cite[Sec.7]{gmpt3}
\begin{itemize}
    \item There exists an \stt\ structure $\phi_\pm$ on $M_6$.
    Here, $\phi_\pm$ are polyforms which are pure spinors for Clifford$(6,6)$, and which satisfy
\begin{equation}\label{eq:stt}
    (\phi_+,\bar\phi_+)=(\phi_-,\bar \phi_-)
    \ ,\qquad
    (\phi_+, X \cdot \phi_-)=0=(\phi_+, X\cdot \bar\phi_-)
\end{equation}
for any $X \in $ \tts. We have used the Chevalley internal product
between internal forms: $(A,A')\equiv (A\wedge A')_6$,
$\lambda(A)\equiv (-1)^{{\rm Int}({\rm deg}(A))}A$.
    \item
    There exist a closed three--form H, an even--degree polyform $F = \sum_k F_{2k}$ (the sum of all the {\it internal} fluxes)
\begin{equation}\label{eq:psp}
    d_H \phi_+ = -2 \mu\, e^{-A} {\rm Re} \phi_-
    \ ,\qquad
    d_H(e^A {\rm Im}  \phi_-)= -3 \mu \,{\rm Im} \phi_+
    + * e^{4A}\lambda (F)
    \ ,\qquad d_H F= 0 \ ,
\end{equation}
where $\Lambda= -3 \mu^2$ is the cosmological constant, and
$d_H \equiv (d - H\wedge)$.
The last equation is actually the Bianchi identity, which can be generalized to contain $\delta$--function--like sources (something we will not do in this paper).
\end{itemize}

If these equations can be solved, $\phi_\pm$ determine a metric $g$, a $b$--field, a dilaton $\phi$ and two six--dimensional  Weyl spinors $\eta^{1,2}_+$. The formulas for the metric and $b$--field in terms of $\phi_\pm$ are a bit involved in general \cite{gualtieri}, but we will see in section \ref{sub:alg} what they give for the cases that we are interested in. The dilaton $\phi$ is determined by
\begin{equation}\label{eq:dil}
    e^{6A-2 \phi} {\rm vol}_6= (\phi_+,\bar \phi_+)
\end{equation}
where vol$_6$ is the volume form determined by the metric $g$; notice that this is {\it not} an extra equation to solve; rather, it determines the dilaton once the supersymmetry equations have been solved. The spinors are determined by
\begin{equation}\label{eq:cl}
    \phi_\pm = e^{-b\wedge}\eta^1_+\otimes \eta^{2\,\dagger}_\pm\ ,
\end{equation}
where $\eta^{1,2}_-\equiv (\eta^{1,2}_+)^*$, and we are confusing a differential form with associated bispinor; one can show
\cite[Sec.~3]{gmpt3} that one can find $\eta^{1,2}_+$ such that (\ref{eq:cl}) is true for any \stt\ structure $\phi_\pm$.
(This fact is crucial in showing that the conditions (\ref{eq:stt}),(\ref{eq:psp}) above are equivalent to the original fermionic equations for supersymmetry.)

We will call the $b$--field determined by $\phi_\pm$ ``intrinsic''. A slight imprecision in (\ref{eq:psp}) is that only if this intrinsic $b$ vanishes, the $*$ in (\ref{eq:psp}) is the usual Hodge star.
This is not a big problem, because one can always obtain a pure spinor pair with vanishing intrinsic $b$ by the action $\phi_\pm \to
e^b \phi_\pm$.

Moreover, there is also an alternative, equivalent formulation of (\ref{eq:psp}) in which $*$ does not appear at all.
It was found in \cite{t-reform}; here we write a more practical version:
\begin{equation}\label{eq:pspb}
    d_{H_0} \phi_+ = -2 \mu\, e^{-A} {\rm Re} \phi_-
    \ ,\qquad
    {\cal J}_+\cdot d_{H_0}(e^{-3A} {\rm Im}  \phi_-)=
    -5 \mu e^{-4A} \,{\rm Re} \phi_+
    +  F^0
    \ ,\qquad d_{H_0} F^0= 0 \ .
\end{equation}
Here ${\cal J}_+\cdot$ is an operator that depends on $\phi_+$
alone; it is explained at length in \cite{t-reform}. In some
cases, its action is easier to compute
than the whole Hodge star. The reason for the appearance of
a subscript ${}_0$ on $H$  in (\ref{eq:pspb}) is that the
 physical $H$ also receives contribution from the $b$ determined
via (\ref{eq:cl}):
\begin{equation}\label{eq:Hph}
    H = H_0 + b\ .
\end{equation}
Notice, however, that it is not necessary to compute $b$ in
order to solve the equations (\ref{eq:pspb}).
Similarly, the physical RR fields are
\begin{equation}\label{eq:Fph}
    F = e^b F^0
\end{equation}
which obey
\begin{equation}
    d_H F = 0 \ .
\end{equation}

In this paper, we will actually be looking for solutions with {\it extended} supersymmetry, namely ${\cal N}=2$ and ${\cal N}=3$.
This simply means that there should be an SO$({\cal N})$ worth of \stt\ structures, all obeying (\ref{eq:psp}) (or (\ref{eq:pspb})) with the same physical fields: the metric $g$, the dilaton $\phi$ and the fluxes $H$, $F$. We will see concretely how this works in section \ref{sec:pert}.


\subsection{Solving the algebraic constraints} 
\label{sub:alg}

We will now analyze the algebraic part of the supersymmetry equations,
(\ref{eq:stt}).

In full generality, there are three cases to consider. Let us call
the {\it type} of a pure spinor $\phi= \sum_{k\ge k_0} \phi_k$ the smallest degree $k_0$ that appears in the sum; in other words, $\phi$
only contains forms of degree type$(\phi)$ or higher. It turns out that the type of a pure spinor in dimension 6 can be at most 3.
There are then three cases:
\begin{enumerate}
    \item $\phi_+$ has type 0, and $\phi_-$ has type 3. This is usually referred to as the ``\sut\ structure'' case, for reasons that will become clear soon.
    \item $\phi_+$ has type 0, and $\phi_-$ has type 1. This is the most generic case, and for this reason it is sometimes just called ``\stt'', or also ``intermediate SU(2) structure''.
    \item $\phi_+$ has type 2, and $\phi_+$ has type 1. This is called ``static SU(2) structure'' case.
\end{enumerate}

In this paper, we are considering small deformations of a solution of type \sut. This will fall in the second, generic \stt, case.
Hence we will now review briefly the solution of the algebraic constraint in the SU(3) structure case, then move on to the generic case, which is our real interest; and we will not pay any attention at all to the static SU(2) structure case.

In the SU(3) structure case, the condition of purity on each $\phi_\pm$ separately determines (up to a $b$--transform)
\begin{equation}\label{eq:su3}
    \phi_+= \rho e^{i \theta} e^{-iJ} \ ,\qquad \phi_- = \rho \,\Omega
\end{equation}
with $\rho$ a complex function, $J$ a non--degenerate two--form, and $\Omega$ a decomposable three--form (one that can be locally written as wedge of three one--forms) such that $\Omega\wedge \bar\Omega$ is never zero\footnote{We are including $(\phi,\bar \phi)\neq 0$ in the definition of purity.}.
The constraint (\ref{eq:stt}) then reduces easily to
\begin{equation}\label{eq:compsu3}
    J \wedge \Omega = 0 \ ,\qquad
    J^3= \frac 34 i \Omega\wedge \bar \Omega \ .
\end{equation}
These equations define an SU(3) structure, which justifies the name given earlier to case 1. We mentioned after (\ref{eq:cl}) that any pair determines an ``intrinsic'' $b$; in this case it is zero. It is more laborious, but also possible, to see that any \stt\ structure determines a metric \cite{gualtieri,gmpt3}. In this case, this works as follows. $\Omega$, being decomposable, determines an almost complex structure $I$ (it is the one such that $\Omega$ is a $(3,0)$--form).
Then we can just define the metric as $g= J I$. The condition (\ref{eq:compsu3}) implies that the $g$ defined in this way is symmetric.

We now come to the case of interest in this paper, namely case 2.
To find the solution to this constraint, one can use \cite{jeschek-witt,minasian-petrini-zaffaroni, halmagyi-t} two different internal spinors $\eta^1_+ \neq \eta^2_+$ in (\ref{eq:cl}); as we remarked earlier, any solution of (\ref{eq:stt}) can be written as in (\ref{eq:cl}), so there is no loss of generality in proceeding this way. One can also \cite{halmagyi-t} solve directly the constraints (\ref{eq:stt}). Either way, one gets
\begin{subequations}\label{eq:nicepair}
    \begin{align}\label{eq:nice+}
        \phi_+ = &\rho\, e^{i \theta}
        \exp \left[-\frac i{\cos (\psi)} j +
        \frac1{2 \tan^2 (\psi)} v\wedge \bar v \right]\ ,\\
        \label{eq:nice-}
        \phi_- = &\rho\, v\wedge \exp \left[\frac 1{\sin (\psi)}
        \left(i{\rm Re} \omega - \frac1{\cos (\psi)} {\rm Im} \omega
        \right) \right]\ ,
    \end{align}
\end{subequations}
for some (varying) angle $\psi$, real function $\rho$, one--form $v$ and two--forms $\omega, j$ satisfying
\begin{equation}\label{eq:SU2}
    \omega^2=0 \ ,\qquad \omega \wedge \bar \omega = 2 j^2 \ ,
\end{equation}
which mean that $\omega, j$ define an SU(2) structure.
Actually, from the constraint (\ref{eq:stt}), one would get (\ref{eq:SU2}) wedged with $v\wedge \bar v$, but one can show \cite[Sec.~3.2]{halmagyi-t} that these can be dropped without any loss of generality.
The pair (\ref{eq:nicepair}) has a non--zero intrinsic $b$--field
(the one defined by (\ref{eq:cl})):
\begin{equation}\label{eq:b}
    b= \tan (\psi) {\rm Im}  \omega\ .
\end{equation}
Notice the difference with the SU(3) structure case, (\ref{eq:su3});
there, the $b$--field of the pair is zero when the exponent of $\phi_+$ is purely imaginary. For (\ref{eq:nicepair}), the exponent of $\phi_+$ is purely imaginary, but the $b$--field is non--vanishing and is given by (\ref{eq:b}).
As we mentioned above, an \stt\ structure also defines a metric. In this case, we get
\begin{equation}\label{eq:metric}
    ds^2 = -j ({\rm Im} \omega)^{-1} {\rm Re} \omega +
    \frac1{\tan^2 (\psi)} v \bar v \ .
\end{equation}
Finally, from the equation (\ref{eq:dil}), we see that the
dilaton $\phi$ is determined by
\begin{equation}\label{eq:rho}
    e^{\phi} = \frac{e^{3A}}{\rho}
\end{equation}
for both cases considered in this subsection, (\ref{eq:su3}) and (\ref{eq:nicepair}).


\subsection{The differential conditions} 
\label{sub:manip}

In this subsection we will take a first look at the differential equations for supersymmetry (\ref{eq:psp}), both for the SU(3) structure case and for the general case.

The SU(3) structure case has been analyzed in \cite{lust-tsimpis}. One can also derive the same conditions from (\ref{eq:psp}) \cite{gmpt3} or
(\ref{eq:pspb}). If we plug (\ref{eq:su3}) in (\ref{eq:pspb}), using ${\cal J}_+ =J\wedge -J\llcorner$ (for more details see \cite{gualtieri}), we see immediately that $F_0=5\mu \cos(\theta)e^{-4A}$. In this paper, we want to perturb SU(3) structure solutions with $F_0=0$ into \stt\ structure solutions with $F_0\neq 0$.
Hence, we only need to give the differential equations for the SU(3) structure case when $F_0=0$. For that reason, we take the angle in
(\ref{eq:su3}) to be
\begin{equation}\label{eq:pi2}
    \theta=\pi/2 \qquad  ({\rm SU}(3)) \ ,
\end{equation}
and we obtain
\begin{equation}\label{eq:su3diff}
\begin{split}
    &d(3A-\phi)=0 \ ,\qquad
    dJ = -2 \mu e^{-A} {\rm Re} \Omega \ ; \\
    &   F_2 = -J^{-1}\llcorner d(e^{-3A}{\rm Im} \Omega)
    +5 \mu e^{-4A} J
    \ ,\qquad F_6 = \frac12 \mu e^{-4A } J^3\ ,
\end{split}
\end{equation}
with $H=F_0=F_4=0$. One could also obtain these equations from M--theory. Notice that\footnote{We thank D.~Martelli and J.~Sparks for discussions on this point.} nothing prevents at this point the warping $A$ (and hence the dilaton $\phi$) from being non--constant, in contrast to the case $F_0\neq0$, in which constancy of $F_0$
(because of its Bianchi identity) implies constancy of $A$. Even though the procedure we outline later for first--order deformations does not require the warping $A$ of the undeformed SU(3) structure solution to be constant, it will be so for the explicit examples of section \ref{sec:cp3}.

We will now look at the \stt\ structure case.  We will actually only solve the supersymmetry equations at first order in perturbation theory; a full analysis of the system (\ref{eq:psp}) in the \stt\ structure case is not really necessary.
Even so, the study of the \stt\ structure case is of independent interest; not many attempts have been made so far for negative cosmological constant (for a recent study, using a particular ``singlet Ansatz'', see \cite{lust-tsimpis-su2singlet}).  We collect here some of the relevant formulas.

We will first look at the first equation in (\ref{eq:psp}) or (\ref{eq:pspb}), and substitute the expression (\ref{eq:nicepair})
for the pure spinors.

The one--form part says that
\begin{equation}\label{eq:rev}
    \rho= \frac1{\sin(\theta)}\ ,\qquad
    {\rm Re} v = \frac{e^A}{2 \mu \sin(\theta)} d \theta \ .
\end{equation}
The three--form part gives, remembering that we choose $\alpha$ to be purely imaginary:
\begin{align}\label{eq:H}
    H_0&= -d (\cot(\theta) J_\psi) \\
    \label{eq:da}
    d\,\left(\frac1{\sin(\theta)} J_\psi \right) &=
    \frac{ 2 \mu e^{-A}}{\sin(\psi)}\left(
    {\rm Im}v\wedge {\rm Re} \omega + {\rm Re} v\wedge
    \frac{{\rm Im} \omega}{\cos(\psi)}\right)\ .
\end{align}
where we have introduced
\begin{equation}\label{eq:Jpsi}
    J_\psi \equiv \frac{j}{\cos(\psi)}+
    i\frac{v\wedge \bar v}{2 \tan^2(\psi)}
\end{equation}
which is none else than $i$ times the exponent of (\ref{eq:nice+}).
Finally, the five--form part can be shown to follow from the
one-- and three--form parts, (\ref{eq:rev}) and (\ref{eq:da}).

Equation (\ref{eq:H}) suggests that we define
\begin{equation}
    \label{eq:B0}
    B_0 =-\cot(\theta) J_\psi\ ,
\end{equation}
which is such that $H_0 = d B_0$.
 We have to remember, however, that the physical $B$--field
also contains another contribution, as we saw in (\ref{eq:Hph})
and (\ref{eq:b}). Hence we get
\begin{equation}\label{eq:btot}
    B = -\cot(\theta) J_\psi
    + \tan(\psi) {\rm Im} \omega \ ,
\end{equation}
up to closed two--forms.

As for the second (and third) equation in (\ref{eq:psp}) or
(\ref{eq:pspb}), we will look at them directly in perturbation theory, since the expressions we obtained are lengthy and not particularly illuminating. The only flux that appears to have a reasonably compact expression is $F_0$. Using the formula for ${\cal J}_+$
relevant for the pure spinor given in (\ref{eq:nice+}),
\begin{equation}\label{eq:Jdot}
    {\cal J}_+ =J_\psi\wedge -J_\psi^{-1}\llcorner\ ,
\end{equation}
after some manipulations we compute
\begin{equation}\label{eq:F0}
    F_0 = -J_\psi^{-1}\llcorner
    d (\rho e^{-3A}{\rm Im} v)
    + 5 \mu e^{-4A}\cot(\theta) \ .
\end{equation}
The expressions for the other fluxes are more conveniently extracted directly from (\ref{eq:psp}). Again, we will see them explicitly in perturbation theory later.


\subsection{Superpotential for D2 probes} 
\label{sub:supot}

We remarked in section \ref{sec:ft} that the abelian version of the superpotential (\ref{eq:w}) should be reproduced by a D2 domain wall, pointlike in the internal manifold $M_6$ and at fixed radius in Poincar\'e coordinates. In this subsection, we compute this superpotential in terms of pure spinors, in a way similar to \cite{martucci} for four--dimensional theories, and anticipated in \cite{koerber-martucci-ads} for thre--dimensional theories. The result will be essential later, in section \ref{sec:pert}, when we will outline the procedure to find infinitesimal perturbations of solutions with no Romans mass.

In massive IIA, let us start with a metric of the form
\begin{equation}\label{eq:37}
    ds^2_{10} = e^{2A_7}ds^2_{{\rm Mink}_{2,1}}+ ds^2_7\ ,
\end{equation}
where the warping factor $A_7$ is a function of the seven internal
coordinates, and the internal metric $ds^2_7$ is so far unrestricted.
We will use the internal fluxes $F$ as an electric basis; they
determine the external fluxes (with legs in the spacetime) via
\begin{equation}
    F_{(10)}= F + {\rm vol}_3 *_7 F \ .
\end{equation}
One can get the equations for ${\cal N}=1$ supersymmetry with a computation similar to the one in \cite{gmpt2}. These equations were considered in \cite{jeschek-witt-shame} in the case without the warping, in \cite[App.~B]{koerber-martucci-ads} for the AdS$_4$ case (which is the one we need here), and they will be presented in general in \cite{haack-lust-martucci-t}.
For our present purposes, we only need to know that they include
\begin{equation}\label{eq:7dpsp}
    d_H(\psi_-)=0\ ,
\end{equation}
where $d_H= d+ H\wedge$, and
$\psi_-$ is part (along with a $\psi_+$ of no relevance here) of a ``generalized G$_2$ structure'' \cite{witt}.

To specialize the equations (\ref{eq:7dpsp}) to a spacetime of the form AdS$_4\times M_6$, we take
\begin{equation}
    ds^2_7 = \frac{e^{2A}}{\mu^2}\frac{d r^2}{r^2} + ds^2_6 \ ,\qquad
    e^{A_7} = \frac{e^A}{\mu} r \ ,
\end{equation}
where $A$ is the warping from the four--dimensional point of view
(the one introduced in section \ref{sec:susy}).
We then decompose
\begin{equation}\label{eq:psiphi}
    \psi_- = \frac{r^2}{\mu^2} \Big(-e^{-A}{\rm Re} \phi_- + \frac{dr}{r\mu} \wedge {\rm Re} \phi_+\Big) \ .
\end{equation}
With this identification, (\ref{eq:7dpsp}) reproduces the real part of the first equation in (\ref{eq:psp}). The rest of (\ref{eq:psp}) can be reproduced too, but we do not need it here.

Now, let us consider a brane that extends along the three external
dimensions, and an internal cycle $B$.
Such a brane is supersymmetric if and only if
\begin{equation}
    (X \cdot \psi_-)_| = 0 \qquad \forall X \in T \oplus T^* \ ,
\end{equation}
where $_|$ denotes pullback to the $B$.
This then suggests that the ${\cal N}=1$ superpotential is
\begin{equation}
    W_{{\cal N}=1} \propto \int_C \psi_- \ ,\qquad \del C = B \ .
\end{equation}
Notice that this makes sense precisely because $\psi_-$ is closed,
(\ref{eq:7dpsp}).
If we now consider the case in which $B$ is a point, we get that
\begin{equation}	
\begin{split}
	d_7 W_{{\cal N}=1} \propto \psi_1 =& -\frac{r^2}{\mu^2}e^{-A} {\rm Re} \phi_1 +
    \frac1{\mu^3} r dr \wedge {\rm Re} \phi_0 \\
    =&\frac1{\mu^3}\left( \frac{r^2}2 d_6 {\rm Re} \phi_0 + r dr\wedge {\rm Re} \phi_0 \right)
    = \frac1{2 \mu^3} d_7 (r^2 {\rm Re} \phi_0)\ .
\end{split}	
\end{equation}
Using (\ref{eq:nice+}) and the first equation in (\ref{eq:rev}), we conclude
\begin{equation}\label{eq:Wtheta}
    W_{{\cal N}=1} = T \cot(\theta)\
\end{equation}
for some proportionality constant $T$.
As we stressed earlier, this D2 superpotential should match the superpotential of the abelianized theory.
Hence the function $\cot(\theta)$, which is one of the data of a gravity solution, is proportional to the abelianized superpotential. This fact should be true for a full solution; but it will be most useful in perturbation theory, as we will now see.



\section{The first--order procedure} 
\label{sec:pert}

We will illustrate here how to start from an SU(3)--structure supersymmetric solution with $F_0=0$, and  perturb it to a first--order  \stt\ solution with $F_0\neq 0$. In section \ref{sub:su3stt} we will explain how to do so at the algebraic level (namely, as far as the constraints in (\ref{eq:stt}) are concerned), and in section \ref{sub:diff} how to solve the differential equations.

\subsection{Perturbing ${\rm SU}(3)$ structure in \stt\ structure} 
\label{sub:su3stt}

The general form of the pure spinors for the SU(3) structure case and for the generic \stt\ structure case have been given in (\ref{eq:su3}) and (\ref{eq:nicepair}).
We will now explain how to take a limit that sends one into the other.

The first thing we want to do is to send the one--form $v$ in (\ref{eq:nice-}) to zero, since $\phi_-$ in (\ref{eq:su3}) has no one--form part. Calling $m$ our first--order deformation parameter, we can write that as
\begin{equation}\label{eq:vm}
    v= m\, v_0 + O(m^2)\ .
\end{equation}
 This creates two potential problems. First,
in the exponent of $\phi_+$ in (\ref{eq:nice+}), we see that the second term would seem to go to zero in the limit $m\to 0$. But (\ref{eq:SU2}) implies $j^3=0$, which means that $j$ is degenerate; since $J$ should be non--degenerate, we should not let the term $v\wedge \bar v/2\tan^2(\psi)$ in (\ref{eq:nice+}) go to zero. This is accomplished by having $\psi$ start its expansion in $m$ at first order:
\begin{equation}\label{eq:psim}
    \psi\to m \psi_0+O(m^2) \ .
\end{equation}
We should also remember that, in the SU(3) structure case, we took $\theta=\pi/2$
(see (\ref{eq:pi2})); hence, we should take
\begin{equation}\label{eq:thm}
    \theta= \frac{\pi}2 + m \theta_0 + O(m^2)\ .
\end{equation}
From the first equation in (\ref{eq:rev}) we also see that $\rho=1+O(m^2)$. Summing up, for $\phi_+$ we get
\begin{equation}\label{eq:phi+m}
    \phi_+ = (i-m \theta_0)\, e^{-iJ}+ O(m^2) \ ,
\end{equation}
with
\begin{equation}\label{eq:Jj}
    J= j + \frac i2 v\wedge \bar v\ ,
\end{equation}
which is a $\psi=0$ limit of (\ref{eq:Jpsi}).
The choice (\ref{eq:psim}) also fixes the second problem created by (\ref{eq:vm}): that it would have risked sending to zero the entire $\phi_-$ in (\ref{eq:nice-}). One might think that now the five--form part will start with a term of order $m^{-1}$ because of the $\frac1{\sin(\psi)}$
in the exponent, but that term is $-\frac 1{2m}v_0 \omega^2 $, which vanishes thanks to (\ref{eq:SU2}). The next term in the expansion is order $m$, and vanishes in the $m\to 0$ limit. The expansion of
$\phi_-$ hence reads
\begin{equation}\label{eq:phi-m}
    \begin{split}
        \phi_- =& v \exp \left[\frac1{m \psi_0}
        \left(i \omega - \frac{m^2 \psi_0^2}2 {\rm Im} \omega \right)\right]+ O(m^2)=\\
        &\left( \frac i{\psi_0} v_0\wedge \omega \right)
        +m v_0 \wedge \left( 1+ \frac 12 \,j^2 \right) + O(m^2)\ .
    \end{split}
\end{equation}
In particular, at order $m^0$, we get
\begin{equation}\label{eq:vom}
    \Omega = \frac i {\psi_0} v_0\wedge \omega\ .
\end{equation}
Hence $v_0$ is a (1,0)--form and $\omega_0$ is a (2,0)--form with respect to the SU(3) almost complex structure $I$ defined by $\Omega$. This is consistent with the constraint $\omega^2=0$ in (\ref{eq:SU2}).


\subsection{Strategy to solve the differential equations} 
\label{sub:diff}

We now move on to the differential equations for supersymmetry, (\ref{eq:psp}). All the equations in this section and in the ones that will follow are to be understood up to orders $O(m^2)$, since we will only solve the equations at first order in perturbation theory.

We begin by noticing that, in the parameterization (\ref{eq:stt})
of the \stt\ structure that we are using, it is natural to divide the various forms according to their parity under reversal of the angle $\psi$. The parity transformations of the pure spinors are
\begin{equation}
    \phi_+ \to - \lambda(\bar \phi_+)
    \ ,\qquad
    \phi_- \to - \lambda( \phi_-) \qquad
     (\psi\to - \psi)\ ;
\end{equation}
recall that $\lambda$ is multiplication by a sign, defined on a $k$--form to be $\lambda(\alpha_k)= (-1)^{\rm{Int}(\frac k2)} \alpha_k$.
From (\ref{eq:psp}), we also see that then the fluxes $H$, $F$
and the warping $A$ transform as
\begin{equation}\label{eq:Fpsi}
    H\to -H \ ,\qquad F \to - \lambda(F)\ ,\qquad A \to A \qquad
     (\psi\to - \psi) \ .
\end{equation}
We took $\psi$ equal to the perturbation parameter $m$, at first order (Eq.~(\ref{eq:psim})). So at order $m^k$, we can consider only the forms with parity $(-1)^k$.

We can now use the expansions in $m$ for $\phi_\pm$ we obtained in
(\ref{eq:phi+m}) and (\ref{eq:phi-m}) in the differential equations (\ref{eq:psp}). In fact, the first equation was already analyzed beyond perturbation theory in section \ref{sub:manip}, so we can just use (\ref{eq:psim}) and (\ref{eq:thm}) in the equations there. Using the remark above about parity under $\psi \to - \psi$, each of these equations will contribute either to order $m^0$ (in which case it should reproduce one of the equations for the SU(3) structure case, (\ref{eq:su3diff})), or at order $m^1$.

The first equation in (\ref{eq:rev}) is even in $m$. It now simply gives that
$\rho=1$, which reproduces $d(3A-\phi)=0$ of the SU(3) structure case (see (\ref{eq:rho}) and (\ref{eq:su3diff})). The second equation in (\ref{eq:rev}) is odd in $m$, and it gives
\begin{equation}\label{eq:rev0}
    {\rm Re} v= m \,{\rm Re}\, v_0=m \frac{e^A}{2\mu}d \theta_0 \ .
\end{equation}
Next, rather than reading (\ref{eq:H}), we can jump at the equation giving the total $B$--field, which is odd in $m$ and reads at first order
\begin{equation}\label{eq:Bm}
    B= m(\theta_0 J + \psi_0{\rm Im} \omega)\ .
\end{equation}
(\ref{eq:da}) is even in $m$ and, at order $m^0$, it simply gives the second equation in (\ref{eq:su3diff}).

We will now look at the expressions for the RR fluxes (the second equation in (\ref{eq:psp}) or (\ref{eq:pspb})). We know from (\ref{eq:Fpsi}) that the equation for $F_2$ and $F_6$ will simply reproduce, at order $m^0$, the corresponding equations in (\ref{eq:su3diff}), and that they will not change at order $m^1$.
In contrast, $F_0$ and $F_4$ will vanish at order $m^0$, but not at order $m^1$. For $F_0$, we can just use (\ref{eq:F0}):
\begin{equation}\label{eq:F0m}
    F_0 = m \Big( -J^{-1}\llcorner d(e^{3A}{\rm Im} v_0)
    -5 \mu e^{-4A}\theta_0\Big)\ .
\end{equation}
We have not given the all--order formula for $F_4$ in section \ref{sub:manip}. We can compute it now by using (\ref{eq:phi+m})
and (\ref{eq:phi-m}) in (\ref{eq:psp}):
\begin{equation}\label{eq:F4m}
    F_4 = e^{-4A}* \Big( m\, d(e^A {\rm Im}  v_0) + 3 \mu B \Big)\ .
\end{equation}

Finally, let us look at the Bianchi identities (the third in (\ref{eq:psp})). The one for $F_0$ simply says that it is constant. The one for $F_4$ is
\begin{equation}\label{eq:dF4}
    dF_4 = H \wedge F_2 \ ;
\end{equation}
recall that there is a non--vanishing $F_2$ in the SU(3) structure solution that we want to deform, and that $H=dB$ and $F_4$ are given by (\ref{eq:Bm}) and (\ref{eq:F4m}).

Notice that $dF_0=0$ and (\ref{eq:dF4}) are the only differential equations we have seen so far. The others are definitions of the fields provided by the supersymmetry equations. At all orders, there would also be equations on the geometry not involving the flux; but, at first order, we just saw that there is no such equation.

To summarize so far, the equations we have to solve at first order in $m$ are (\ref{eq:dF4}) and that $F_0$ in (\ref{eq:F0m}) is constant. If one wants to have extended supersymmetry, we remarked at the end of section \ref{sub:susy} that one is actually looking for a SO$({\cal N})$ worth of pure spinors, but in such a way that the physical fields (the fluxes, the metric and the dilaton) are invariant.
In that case, one will then have to impose by hand that $B$, $F_0$ and $F_4$ in (\ref{eq:Bm}), (\ref{eq:F0m}) and (\ref{eq:F4m}) are invariant.

We will now see that there is not much freedom in solving these equations: for an assigned field theory, no guesswork is necessary.

First of all we should remember  (\ref{eq:Wtheta}). That equation should be true at all orders, but at first order it just says
\begin{equation}\label{eq:Wth0}
    W_{{\cal N}=1} = -m T \theta_0  \ .
\end{equation}
Now, $v_0$ follows by combining (\ref{eq:rev0}) with the fact that it is a (1,0)--form
with respect to the almost complex structure of the SU(3)--structure solution:
\begin{equation}\label{eq:vW}
    v_0 = \frac{e^A}{\mu} \del \theta_0
\end{equation}
where $\del$ is the Dolbeault operator. We can now find $\omega$ and $j$ from the data of the SU(3) structure, $J$ and $\Omega$. For $\omega$, we can use (\ref{eq:vom}) combined with the $\psi\to0$
limit of (\ref{eq:metric}); for $j$, we can simply invert (\ref{eq:Jj}):
\begin{equation}\label{eq:oj}
    \omega= -\frac i{2 \psi_0} \bar v_0 \llcorner \Omega
    \ ,\qquad
    j= J - \frac i2 v\wedge \bar v\ .
\end{equation}
At this point the fluxes are going to be determined via (\ref{eq:Bm}),
(\ref{eq:F0m}) and (\ref{eq:F4m}); there are no choices to be made. All one has to do is to check that the supersymmetry equations explained earlier hold.
In this sense, our procedure is algorithmic. Once one knows from field theory arguments the right $W_{{\cal N}=1}$, the gravity dual is determined at first order in perturbation theory.

Let us summarize. Suppose one has a CFT$_3$ dual to a supersymmetric SU(3) structure AdS$_4$ vacuum of IIA; most AdS$_4$/CFT$_3$ duals known are of this type. Suppose one identifies a new conformal field theory that deforms the old one, in a way which is dual to switching on a Romans mass; examples of such deformations were given in \cite{gaiotto-t}. If the superpotential of this theory is non--vanishing even at the abelian level, the gravity dual will be a solution of \stt\ structure type, and it will be given, at first order in the Romans mass, by the procedure outlined in the preceding paragraph.

In the next section, we will illustrate this procedure by finding the perturbative solutions dual to the theories reviewed in section \ref{sec:ft}.



\section{Perturbative solutions on ${\rm AdS}_4\times \cc\pp^3$} 
\label{sec:cp3}

In this section, we will apply the procedure outlined in section \ref{sec:pert} to the theories discussed in section \ref{sec:ft}.
We will start by reviewing briefly, in section \ref{sub:n6} and \ref{sub:n6ang}, the SU(3) structure solution we want to deform, in two sets of coordinates convenient to our needs.  In the remaining subsections, we will find the ${\cal N}=2$ and ${\cal N}=3$ gravity duals we promised.

\subsection{The ${\cal N}=6$ solution in homogeneous coordinates} 
\label{sub:n6}

In this section, we will review the ${\cal N}=6$ solution \cite{nilsson-pope,sorokin-tkach-volkov-11-10} on
AdS$_4\times\cc\pp^3$ from the IIA point of view.

When we discussed the differential supersymmetry conditions for
SU(3) structure in section \ref{sub:manip}, we found in equation
(\ref{eq:su3diff}) that $J$ cannot be closed (recall that $\Lambda =
-3 \mu^2$). Hence, it cannot be a \ka\ form, and in particular not
the usual Fubini--Study \ka\ form $J_{\rm FS}$. Also, one could not
even write an  $\Omega_{\rm FS}$ which is globally defined and which
is (3,0) with respect to the usual complex structure on $\cc\pp^3$,
since, for that complex structure, $c_1=4$. Fortunately, there are
other almost complex structures on $\cc\pp^3$, with respect to which
$c_1=0$ (so that a globally defined (3,0)--form $\Omega$ exists),
and so that $J$ is not closed. There is an $S^5$ worth of such almost
complex structures; each point in this $S^5$ corresponds to a supersymmetry of the ${\cal N}=6$ solution.

Let us start from $\cc^4$, with coordinates $z^A$, $A=1,\ldots,4$. One can think of $\cc^4-\{ \underline{0} \}$  as
a $\cc^*$ bundle over $\cc\pp^3$ (with missing zero section), with projection map $p$. A form $\alpha$ on the total space of a bundle with projection $p$ is the pull--back of a form on the base space if and only if it is basic, namely if it is vertical ($\iota_v \alpha=0$, for any $v$ tangent to the fibres of $p$) and invariant (its Lie derivative with respect to any $v$ tangent to the fibres of $p$ vanishes, $L_v \alpha=0$). In our case, the forms
\begin{equation}\label{eq:Dz}
    D z^A = d z^A - z^A \frac{\bar z_B d z^B}{\bar z_C z^C}=
    \left(\delta^A{}_B - \frac{z^A \bar z_B}{\bar z_C z^C}\right) dz^B
    \equiv P^A{}_B dz^B
\end{equation}
are basic: they are annihilated by contraction with both vectors
\begin{equation} \label{eq:rxi}
    r\del_r= z^A \del_A + \bar z_A \del^{\bar A} \ ,\qquad
    \xi= i(z^A \del_A - \bar z_A \del^{\bar A}) \qquad
    (r^2= z^A \bar z_A)\ ,
\end{equation}
and they are closed (which, together with (\ref{eq:rxi}), implies that they are also invariant). Hence, they are pull--back of forms on $\cc\pp^3$.
We will use the projector $P^A{}_B$ in (\ref{eq:Dz}) to do computations on $\cc\pp^3$ using coordinates of $\cc^4$.

Another way of thinking about (\ref{eq:Dz}) is the following: given a form on $\cc^4$, one can try to define a basic form by subtracting its non--vertical part. In terms of the one--forms
\begin{equation}
    rdr= \frac12 (\bar z_A dz^A + z^A d\bar z_A)\ ,\qquad
    \eta = \frac i{2 r^2} (-\bar z_A dz^A + z^A d\bar z_A)\ ,
\end{equation}
in the case of the form $dz^A$, this decomposition reads
\begin{equation}\label{eq:Dzeta}
    d z^A = D z^A + z^A \left( \frac{dr}r + i \eta \right) \ .
\end{equation}
The one--form $\eta$ is dual to $\xi$ above, in that $\iota_\xi
\eta=1$. We can apply the same procedure to the standard \ka\ form
in $\cc^4$:
\begin{equation}\label{eq:jfs}
    J_{(4)}= \frac i2 dz^A\wedge d\bar z_A=
    rdr\wedge \eta + r^2 J_{\rm FS} \ ,\qquad
    J_{\rm FS} = \frac i{2r^2} Dz^A \wedge D\bar z_A\ .
\end{equation}
The explicit expression of $J_{\rm FS}$ on the right makes it clear that it is basic. One can also see that $J_{\rm FS}$ is vertical from its definition on the left, using that $\iota_{r\del_r} J_{(4)}= r^2 \eta$; using the fact that $J_{(4)}$ is quadratic, $L_{r\del_r}J_{(4)}=2 J_{(4)}$, one can also see easily that
\begin{equation}
    d \eta = 2 J_{\rm FS}\ ,
\end{equation}
which implies that $J_{\rm FS}$ is also invariant under $\eta$.
This $J_{\rm FS}$ is the standard Fubini--Study \ka\ form on
$\cc\pp^3$. As we remarked earlier, however, it is not exactly what we need in the supersymmetry equations.

To construct the supersymmetric $J$, we need to introduce more data.
A {\it holomorphic symplectic form} $\kappa$ in four complex dimensions is a two--form whose square gives the holomorphic volume form $\Omega_{(4)}$:
\begin{equation}\label{eq:oO}
    \frac12\kappa^2 = \Omega_{(4)} \ .
\end{equation}
In $\cc^4$, one has an $S^5$--worth of holomorphic symplectic forms
$\kappa= \kappa_{AB} dz^A dz^B$. From each of these, one can extract
the radial and non--radial parts using the vector $r\del_r$ and $\xi$,
just like in (\ref{eq:Dzeta}) and (\ref{eq:jfs}):
\begin{equation}\label{eq:omegadec}
    \kappa= r(dr +i r\eta)\wedge s_\kappa+ r^2 t_\kappa\ ,
\end{equation}
In components, using the forms (\ref{eq:Dz}), one can also write
\begin{equation}\label{eq:st}
    s_\kappa = \frac1{r^2}\kappa_{AB}z^A D z^B \ ,\qquad
    t_\kappa = \frac1{2r^2}\kappa_{AB} Dz^A \wedge Dz^B\ .
\end{equation}
These forms are vertical by construction, but they are not invariant under $\xi$. By comparing (\ref{eq:omegadec}) with $L_{r\del_r}\kappa= 2\kappa$, one obtains
\begin{equation}\label{eq:ds}
    d s_\kappa = 2(i \eta\wedge s_\kappa + t_\kappa)\ ,
\end{equation}
and, from this, $L_\xi s_\kappa= 2i s_\kappa$,  $L_\xi t_\kappa= 2i t_\kappa$. Similarly, if one defines a vertical form $\Omega_{\rm FS}$ by
\begin{equation}
    \Omega_{(4)}= r^3(dr + i r \eta)\wedge \Omega_{\rm FS} \ ,
\end{equation}
one sees that $L_\xi \Omega_{\rm FS}= 4i \Omega_{\rm FS}$ (which is related to the fact that $c_1=4$). In fact, by using our definition (\ref{eq:oO}) above, we get
\begin{equation}
    \Omega_{\rm FS}= s_\kappa \wedge t_\kappa
\end{equation}
for any holomorphic symplectic $\kappa$. So $\Omega_{\rm FS}$ does not define a form on $\cc\pp^3$.

This, however, suggests a way of defining a different three--form which is both vertical and invariant:
\begin{equation}\label{eq:Omega}
    \Omega_\kappa \equiv -i \bar s_\kappa \wedge t_\kappa \ ;
\end{equation}
this time $L_\xi \Omega_\kappa=0$, because the charges of $\bar s_\kappa$ and $t_\kappa$ add up to zero, rather than to $4$ as for $\Omega_{\rm FS}$.
(The factor $-i$ has no particular meaning; it has been selected for consistency of notation with the previous sections.)
The new three--form now defines a new almost complex structure $I$, under which it is a (3,0) form. Roughly speaking, we have just conjugated the usual Fubini--Study complex structure in one direction out of three.\footnote{$\cc\pp^3$ can also be thought of as the twistor space of $S^4$; the new almost complex structure corresponds then to conjugation on the $\cc\pp^1$ fibre. This second almost complex structure makes sense on any twistor space \cite{eells-salamon}.}

For supersymmetry, we need to complement $\Omega_\kappa$ in (\ref{eq:Omega})
with a $J$ that obeys (\ref{eq:compsu3}). If we decompose $J_{\rm FS}$ as
\begin{equation}
    J_{\rm FS}= j_\kappa + \frac i2 s_\kappa \wedge \bar s_\kappa\ ,
\end{equation}
the remark we just made about the new almost complex structure defined by  (\ref{eq:Omega}) suggests that we define
\begin{equation}\label{eq:Jo}
    J_\kappa= j_\kappa- \frac i2 s_\kappa \wedge \bar s_\kappa = J_{\rm FS}
    -i s_\kappa \wedge \bar s_\kappa\ .
\end{equation}
Notice that this form is also well--defined on $\cc\pp^3$, because the term $s_\kappa\wedge\bar s_\kappa$ is invariant under $\eta$. Using now (\ref{eq:ds}) and some manipulations, it is not difficult to see that (\ref{eq:compsu3}) and (\ref{eq:su3diff}) are satisfied by
\begin{equation}\label{eq:n6sol}
    J=J_\kappa \ ,\qquad \Omega= \Omega_\kappa \ ;\qquad
    F_2 = d \eta = 2 J_{\rm FS} \ ,\qquad \mu= -2 \ ,\qquad A=0\ .
\end{equation}
Since this solution works for any holomorphic symplectic form $\kappa$
(see (\ref{eq:oO})), and there is an $S^5$ worth of such forms on $\cc^4$, we conclude that this solution has ${\cal N}=6$.

Before we move on to the perturbative solutions, let us also remark that one can also use homogeneous coordinates to describe the ${\cal N}=1$ massive solutions in \cite{t-cp3}. One simply has to rescale $j$ and $t$ by a factor of $2/\sigma$, so that\footnote{In \cite{t-cp3},
the ${\cal N}=6$ solution is recovered for $\sigma=2$. In this paper, we use slightly different conventions: the ${\cal N}=6$ solution in (\ref{eq:n6sol}) is obtained by again setting $\sigma=2$ in (\ref{eq:n1}), followed by an additional (immaterial) conjugation $J\to -J$,
$\Omega\to \bar \Omega$, $g\to g$. Also, for consistency with \cite{t-cp3} we introduced in (\ref{eq:n1}) the curvature radius $R$, which we have set to one in the rest of this paper.\label{foot:n1}}
\begin{equation}\label{eq:n1}
    J_\sigma = R^2 \left(-\frac 2{\sigma} j + \frac i2 s_\kappa \wedge \bar s_\kappa \right) \ ,\qquad
    \Omega_\sigma = R^3 \frac{2i}{\sigma} s_\kappa\wedge \bar t_\kappa\ .
\end{equation}
The formulas for the fluxes can then be found in \cite[Eq.~(2.2)]{t-cp3}.


\subsection{The $T^{11}$ foliation} 
\label{sub:n6ang}

We present here the ${\cal N}=6$ solution in a different set of coordinates, first used in \cite{cvetic-lu-pope-cpn}, which are adapted to the foliation of $\cc\pp^3$ in $T^{11}= S^2 \times S^3$. These coordinates will allow us to offer, later on, an alternative presentation of one of our solutions, the one with SO(4)$\times {\rm U}(1)_{\rm R}$ isometry group (discussed in section \ref{sub:n2}).

Before we discuss the foliation, let us review some useful forms on $S^2$, that we will then use on each of the $S^2$s in (\ref{eq:t11fol}).
In terms of the usual holomorphic coordinates on $S^2$, $z = \tan
\left(\frac{\theta}{2}\right) e^{i \phi}$, we have the one--form
\begin{equation}
e=\frac{2 dz}{1+|z|^2} = e^{i \phi}(d\theta + i \sin \theta d\phi) \  ;
\end{equation}
the round metric is then $ds^2_{S^2} = e \bar e$, and the K\"ahler
form $J=\frac{i}{2} e \wedge \bar e$. Also,
\begin{equation}
d e = i A\wedge e\ ; \qquad A = i \frac{z d \bar z- \bar z dz}{1+|z|^2}\ .
\end{equation}
In usual coordinates, $A = (1 - \cos \theta) d\phi$; note that $dA = J_{S^2}$.  Of course globally $J_{S^2}$ is not exact (it is the K\"ahler form of $S^2$), and the expressions we just wrote are valid in a patch.
Finally, notice also that $e$ and $\bar e$ are related to the SU(2)--invariant forms $\sigma_i$ on $S^3$ via the Hopf fibration: if one adds an angle $\psi$, one has
\begin{equation}
    \sigma_1 + i \sigma_2 \equiv \sigma_+ = e^{i \psi} \bar e \ ,\qquad
    \sigma_3 = d \psi - A \ .
\end{equation}
These forms satisfy $d \sigma_i = \frac12\epsilon_{ijk}\, \sigma_j \wedge \sigma_k$, as appropriate for left--invariant forms on $S^3$.
Notice also that, in these conventions, the round metric on $S^3$ with radius one is
\begin{equation}
    ds^2_{S^3}= \frac14\sigma_i \sigma_i =\frac14 \left( ds^2_{S^2} +
    (d \psi -A )^2 \right) \ .
\end{equation}
In what follows, we will use the forms $A$, $J$, $e$ we just introduced on each of the $S^2$, with a subscript ${}_i$, $i=1,2$, denoting which of the two $S^2$s it refers to.

We will now discuss the $T^{11}$ foliation of $\cc\pp^3$. From the point of view of the field theory, this foliation exists because of a simple relation \cite{jafferis-t,martelli-sparks-3dquivers,hanany-zaffaroni-cs} between the moduli spaces of the Chern--Simons--matter theory and of the four--dimensional theory with the same quiver, which is in this case the conifold theory \cite{klebanov-witten}.    
On the gravity side, it comes about as follows.  The splitting $\rr^8  =
\rr^4 \times \rr^4$ allows one to realize $S^7$ as a fibration of
$S^3 \times S^3$ on a segment. We can parameterize the segment as an angle $0 \le t \le \pi/2$; the radii of the two $S^3$s are $\cos(t)$
and $\sin(t)$:
\begin{equation}
\begin{split}
    ds^2_{S^7}=&
    dt^2 +\cos^2(t) ds^2_{S^3_1}
    + \sin^2 (t) ds^2_{S^3_2}\\
    =& dt^2 + \frac14 \left(\cos^2 (t) ds^2_{S^2_1} +
    \sin^2 (t) ds^2_{S_2^2}+
    \cos^2 (t) (d\psi_1 - A_1)^2 +
    \sin^2 (t) (d\psi_2 - A_2)^2 \right)
\end{split}
\end{equation}
We can now rearrange $2\psi_1 = \psi + a$  and $2\psi_2 = \psi - a$, and reduce on the angle $\psi$. Each of the leaves at $\{ t=t_0 \}$
gets reduced from $S^3 \times S^3$ to $T^{11}= S^3\times S^2$. The reduction on $\psi$ is nothing but the Hopf fibration to $\cc\pp^3$; hence we have realized $\cc\pp^3$ as a foliation whose generic leaves are copies of $T^{11}$. Even at the level of the metric we can write:
\begin{equation}\label{eq:t11fol}
ds^2_{\cc\pp^3} =
dt^2 + \frac14 \left(\cos^2 (t) ds^2_{S^2_1} +  \sin^2 (t) ds^2_{S_2^2}+
\sin^2 (t) \cos^2 (t)  (Da)^2 \right)\ ,
\end{equation}
where
\begin{equation}
    Da =  da - A_1 + A_2\ ;
\end{equation}
notice that $d(Da) = J_2 - J_1$.
The Fubini--Study K\"ahler form then reads (again in a patch)
\begin{equation}
\begin{split}
    4 J_{\rm FS}&=\frac12 d[A_1 + A_2 - \cos (2t) Da ] \\
    &=\cos^2(t) J_1 + \sin^2 (t) J_2 + \sin (2t) dt\wedge Da \ .
\end{split}
\end{equation}
A simple basis of $(1,0)$--forms for the usual complex structure $I_{\rm FS}$ is $2dt + i \sin (t) \cos (t) Da$, $\cos (t) e_1$,
$\sin (t) e_2$.

It is useful to define also
\begin{equation}
    \omega_- = \frac{i}{2} e^{-i a} e_1 \wedge \bar e_2 \ ,\qquad
    \omega_+ =
    \frac{i}{2} e^{i a}  e_2 \wedge \bar e_1\ ,
\end{equation}
which are SU(2) $\times$ SU(2) invariant. Indeed,
the angle $a$, together with the two $S^2$s, builds up a
$T^{11} =({\rm SU}(2)_1 \times {\rm SU}(2)_2)/{\rm U}(1)$.
On $S^3\times S^3$ we have the left--invariant forms $\sigma_{1,2}^3, \sigma_{1,2}^{\pm}$. The form $\sigma_1^3 + \sigma_2^3$ is zero on the quotient. To make  SU(2)$\times$SU(2)--invariant
forms we are supposed to take combinations of the remaining five
forms, in a way which is invariant under the U(1) action we
quotient by. Among these we find
$\sigma_{1\,\pm} \wedge \sigma_{2\,\mp}= \pm 2i \omega_\mp$. Notice also that $\omega_+ \wedge \omega_- = - J_1 \wedge J_2$, and that $\omega_\pm$ have charge $\pm 1$ under the U(1) isometry $\del_a$; in particular,
$d\omega_\pm =\pm i Da\wedge \omega_\pm$.

So far we have used the usual, integrable complex structure on $\cc\pp^3$. As we saw in section \ref{sub:n6}, however, for IIA supersymmetry we need a different almost complex structure. There, we introduced an $S^5$ worth of SU(3) structures $(J_\kappa, \Omega_\kappa)$
that represent the six supersymmetries of the Fubini--Study metric.
In the coordinates we are using in this section, only two of these SU(3) structures will be manifest (or, to be more precise, a U(1) worth of them). Of course one can write a similar foliation in many different ways, and make manifest the other SU(3) structures which we know to exist. In any case, the ones we will see in the present set of coordinates will be enough to give an alternative presentation of the solution in section \ref{sub:n2}.
These SU(3) structures will be SO(4)--invariant after the infinitesimal deformation of section \ref{sub:n2}, but they actually have SO(5) invariance before the deformation.

This SO(5) invariance is present because of the existence of a fibration $\cc\pp^3 \to S^4$, with fibre $S^2$. We will now describe how this projection is compatible with the $T^{11}$ foliation we just saw. If a point of $T^{11}$ is given as
a pair of SU(2) elements $(g_1,g_2)$ up to the diagonal U(1)
action on the right, there is a natural projection onto an $S^3$:
\begin{equation}
(g_1,g_2) \to g=g_1 g_2^{-1}
\end{equation}
Notice that $dg g^{-1}  = dg_1 g_1^{-1} - g_1 g_2^{-1} dg_2 g_1^{-1}$, so that the round metric on the $S^3$ pulls back to
\begin{equation}
    \begin{split}
    {\rm Tr} (dg g^{-1} dg g^{-1}) = {\rm Tr}  (dg_1 g_1^{-1} dg_1 g_1^{-1}) + & {\rm Tr}  (dg_2 g_2^{-1} dg_2 g_2^{-1}) - 2 {\rm Tr} (g_1^{-1} dg_1 g_2^{-1} dg_2) \\
    &= (\sigma_1 - \sigma_2)_i(\sigma_1 - \sigma_2)_i\ ;
    \end{split}
\end{equation}
the ${}_{1,2}$ here refer to one of the two spheres, whereas the ${}_i$ to one of the three left--invariant forms.
Hence the pullback of the round metric on an $S^4$ of radius 2 is
\begin{equation}
ds^2_{S^4} =
dt^2 + \frac14 \sin^2(2 t) ds^2_{S^3} = dt^2 + \frac1 {16}\sin^2 (2t) ( (Da)^2 +  (\sigma_1 - \sigma_2)_+(\sigma_1 - \sigma_2)_-) \ ;
\end{equation}
recall that $t$ goes from $0$ to $\pi/2$.
If we subtract this from (\ref{eq:t11fol}), we expect to find the metric on the $S^2$ fibre of the fibration $\cc\pp^3 \to S^4$.
We get
\begin{equation}
    \begin{split}
        &\frac14 \left(\cos^2 (t) \sigma_{1\,+} \sigma_{1\,-} +
        \sin^2 (t) \sigma_{2\,+} \sigma_{2\,-} -
\sin^2 (t) \cos^2 (t) (\sigma_1 - \sigma_2)_+(\sigma_1 - \sigma_2)_-
\right)= \\
        & \frac14(\cos^2 (t) \sigma_{1\,+} + \sin^2 (t) \sigma_{2\,+})
        (\cos^2 (t) \sigma_{1\,-} + \sin^2 (t) \sigma_{2\,-})
        \ ,
    \end{split}
\end{equation}
which is indeed of rank $2$. Locally, we can now give three holomorphic vielbeine:
\begin{equation}\label{eq:holviel}
\begin{split}
        E_1 &= dt + \frac i4 \sin (2t) Da \ ,\\
        E_2 &= \frac14\sin (2t) (e^{- i a/2}e_1
    - e^{i a/2} e_2) \ , \\
        E_3 &= \frac12
        \left(\cos^2 (t)  e^{- i a/2} e_1 + \sin^2 (t) e^{i a/2} e_2
        \right)\ ,
\end{split}
\end{equation}
in terms of which
the Fubini--Study K\"ahler form can be written as
\begin{equation}
   J_{\rm FS} =
\frac{i}{2} E_i \wedge \bar E_i \ .
\end{equation}

The almost complex structure appropriate for supersymmetry can now be found by conjugation on the $S^2$ fibre, as in \cite{t-cp3} and as in section \ref{sub:n6}. Namely, we define the $(3,0)$ form $\Omega$ to be
\begin{equation}
\begin{split}
    \Omega =& i E_1 \wedge E_2 \wedge \bar E_3 =\\
    &=\frac14\sin (2t) \left(dt + \frac i4 \sin (2t) Da\right)\wedge
    \left(\cos^2 (t) J_1 - \sin^2 (t) J_2 - \cos^2 (t) \omega_+ + \sin^2 (t) \omega_-\right)\ .
\end{split}
\end{equation}
The two--form $J$ is then determined to be, if ones does not wish to modify the metric:
\begin{equation}
\begin{split}
    J = & J_{\rm FS} - i E_3 \wedge \bar E_3 =
    \frac{i}{2} \left( E_1\wedge \bar E_1 + E_2\wedge \bar E_2 + \bar E_3 \wedge E_3\right)\\
    =&\frac14 \left( \sin (2t) \,dt\wedge Da - \cos (2 t) \cos^2 (t) J_1 + \cos (2 t) \sin^2 (t) J_2- \frac12 \sin^2 (2t) (\omega_+  + \omega_-) \right)\ .
\end{split}
\end{equation}

We conclude this section by remarking that these coordinates can also be used, as could the ones we saw in section \ref{sub:n6}, to reproduce the ${\cal N}=1$ solutions in \cite{t-cp3}. The SU(3) structure data read
\begin{equation}
    J_\sigma = \frac i2 R^2 \left( -\frac 2 \sigma (E_1 \wedge \bar E_1
    + E_2 \wedge \bar E_2) + E_3 \wedge \bar E_3  \right)
    \ ,\qquad
    \Omega_\sigma = -\frac {2i} \sigma R^3 \bar E_1 \wedge \bar E_2
    \wedge E_3 \ ;
\end{equation}
see also the comments in footnote \ref{foot:n1}.


\subsection{The ${\cal N}=2$ solution with enhanced isometry group} 
\label{sub:n2}

In this subsection, we will apply the procedure of section \ref{sec:pert} in detail to one of the field theories in section \ref{sec:ft}, namely the one for which there is enhanced SU(2)$\times$SU(2)($\times$ SO(2)$_{\rm R}$) global symmetry; this theory corresponds to some $c_i$ on the locus (\ref{eq:n2enh}). We gave the ${\cal N}=2$ non--abelian superpotential for this theory in (\ref{eq:w}); the superpotential we need in (\ref{eq:vW}) is the
abelian ${\cal N}=1$ superpotential. We have to rewrite the the theory
in (\ref{eq:cft2}) in terms of ${\cal N}=1$ superfields; there will be a term, then, of the form $\int d^2 \theta W_{{\cal N}=1}$, and this $W_{{\cal N}=1}$ is the one we need to abelianize. It is a real function, with a contribution $\sim {\rm Re} (W_{{\cal N}=2})$
and a contribution from D--term couplings. The result can be conveniently expressed in terms of two of the constant holomorphic symplectic forms we defined in section \ref{sub:n6}:
\begin{equation}\label{eq:W1n2enh}
    W_{{\cal N}=1}=
    \left(\frac{2 \pi}{k_1} + \frac{2 \pi}{k_2} \right) \frac{\nu^2}2 \ ,\qquad \nu =
    i\,\frac1{r^2}\bar z_A \kappa^{AB} \tilde\kappa_{BC} z^C\ .
\end{equation}
Here $\kappa^{AB}\equiv \epsilon^{ABCD}\kappa_{CD}$, and via a change of coordinates (see footnote \ref{foot:chb}) we take $\kappa$ and $\tilde \kappa$ to anticommute, so that $\kappa\tilde \kappa$ is antisymmetric; the $i$ in the definition of $\nu$, then, makes sure it is real. We know from (\ref{eq:Wth0}) that (\ref{eq:W1n2enh}) is proportional to $\theta_0$; at this point we have not specified what the parameter $m$ is, and we can fix it by the choice
\begin{equation}\label{eq:th0enh}
    \theta_0 = \frac12 \nu^2\ ,
\end{equation}
that will be convenient later.

Recall from section \ref{sub:n6} that each of the six supersymmetries corresponds to  an SU(3) structure $(J_\kappa,\Omega_\kappa)$ associated to a holomorphic symplectic form $\kappa$ via the formulas (\ref{eq:Jo}),
(\ref{eq:Omega}). Out of those six SU(3) structures, we are only interested in the two associated to the holomorphic symplectic forms $\kappa$, $\tilde \kappa$ appearing in (\ref{eq:W1n2enh}); those are the two SU(3) structures that we want to deform into \stt\ structure solutions. In the following, we will deform the SU(3) structure associated to $\kappa$; we will check at the end that one could have used $\tilde\kappa$ and obtained another solution to the supersymmetry equations with the same flux. These two solutions, then, are actually a single ${\cal N}=2$ solution, as we explained at the end of section \ref{sub:susy}.

We can now turn the crank of the machine described at the end of section \ref{sub:diff}, specialized to the ${\cal N}=6$ solution described in section \ref{sub:n6}. First of all,  (\ref{eq:vW}) instructs us to take the Dolbeault derivative of (\ref{eq:W1n2enh}). This should be done with respect to the almost complex structure $I_\kappa$ associated to $\Omega_\kappa$. For the particular superpotential in (\ref{eq:W1n2enh}), there is a simplification. Compute the Dolbeault derivative with respect to the Fubini--Study complex structure,
\begin{equation}\label{eq:delFS}
    \del_{\rm FS} \theta_0 =
    i\frac{\nu}{r^2}\,\bar z_A \kappa^{AB} \tilde\kappa_{BC} Dz^C \ ,
\end{equation}
where $Dz^A$ was defined in (\ref{eq:Dz}).
Using (\ref{eq:delFS}), one actually finds that
\begin{equation}
    \bar s_\kappa \llcorner \del_{\rm FS} \theta_0 = 0\ ;
\end{equation}
this means that $\del_{FS} \theta_0 $ does not have any component along $s_\kappa$, which is the direction in which $I_\kappa$ and $I_{\rm FS}$ differ by conjugation. This means, then, that
\begin{equation}
    v_0= -\frac12
    \del \theta_0 =-\frac 12 \del_{\rm FS}  \theta_0 =
    -\frac{i\nu}{2 r^2}
    \,\bar z_A \kappa^{AB} \tilde\kappa_{BC} Dz^C\ ;
\end{equation}
we have used $A=0$, $\mu=-2$, coming from (\ref{eq:n6sol}).

We can now also compute, using (\ref{eq:oj}):
\begin{equation}
    \omega= \frac{\nu}{2 \psi_0} \bar s_\kappa \wedge
    (i s_{\tilde \kappa}+ \nu s_\kappa)\ .
\end{equation}
Using (\ref{eq:Bm}) we can now see that
\begin{equation}\label{eq:Bn2enh}
    B= m \frac \nu 2  ( \nu J_{\rm FS}
    + {\rm Re} (\bar s_\kappa \wedge s_{\tilde \kappa}) )\ .
\end{equation}
We then find
\begin{equation}
    d {\rm Im}  v_0 = -2 \frac i {\nu^2} v_0 \wedge \bar v_0
    - \frac \nu {2 r^2} D\bar z_A\wedge \kappa^{AB}\tilde \kappa_{BC} Dz^C
    + \nu^2 J_{\rm FS}  \ ;
\end{equation}
using this in (\ref{eq:F0m}), and knowing from (\ref{eq:Jo})
that
\begin{equation}
    J_\kappa^{-1}=J_{\rm FS}^{-1}-i (J_{\rm FS}^{-1} s_\kappa)
    \wedge (J_{\rm FS}^{-1} \bar s_\kappa)\ ,
\end{equation}
we get
\begin{equation}\label{eq:F0eqm}
    F_0 = m\ .
\end{equation}
in fact, in (\ref{eq:th0enh})
we adjusted our choice of proportionality constant between $W_{{\cal N}=1}$ and $\theta_0$ so as to get exactly (\ref{eq:F0eqm}). Notice that comparison between (\ref{eq:W1n2enh}), (\ref{eq:th0enh}), (\ref{eq:Wth0}) and (\ref{eq:F0k}) now gives $T=-\frac{2 \pi}{k_1 k_2 }\sim\frac{2\pi}{k^2}$. It would be interesting to compute this more directly using the brane probe logic of section \ref{sub:supot}.

Finally, we can compute from (\ref{eq:F4m})
\begin{equation}\label{eq:F4n2enh}
    * F_4= m \left[-2 \frac i {\nu^2} v_0 \wedge \bar v_0
    - \frac \nu {2 r^2} D\bar z_A\wedge \kappa^{AB}\tilde \kappa_{BC} Dz^C -2 \nu^2 J_{\rm FS} -3 \nu {\rm Re} ( \bar s_\kappa \wedge
    s_{\tilde \kappa})
    \right]
\end{equation}

We can now check whether these are the data of a solution or not. $F_0$ is manifestly a constant; it is a bit more involved to check that $F_4$ satisfies indeed (\ref{eq:dF4}), with  $F_2=J_{\rm FS}$, as in the ${\cal N}=6$ solution (see (\ref{eq:n6sol})), and with $H=dB$, $B$ being given in (\ref{eq:Bn2enh}). So this is a supersymmetric solution. We now ask whether it is an ${\cal N}=2$ solution. The computation so far consisted in perturbing the SU(3) structure $(J_\kappa, \Omega_\kappa)$; we now have to consider what happens if we exchange the roles of $\kappa$ and $\tilde \kappa$. As we remarked earlier, it is enough to check whether the physical fields are invariant under such an exchange; this is manifestly true for $B$ and $F_4$ in (\ref{eq:Bn2enh}) and
(\ref{eq:F4n2enh}). As one expects, there is also an R--symmetry U(1)$_{\rm R}$ that rotates $\kappa$ and $\tilde \kappa$, and that leaves the fluxes invariant.

Hence we have found an ${\cal N}=2$ solution on
AdS$_4\times \cc\pp^3$ with \stt\ structure, as a perturbation
of the ${\cal N}=6$ solution in section \ref{sub:n6}.

\subsubsection{An alternative presentation using the $T^{11}$ foliation} 
\label{subsub:alt}

We can rewrite the solution we just found, in the coordinates we worked out in section \ref{sub:n6ang}. First of all, $\theta_0$ can be written as
\begin{equation}
    \theta_0 = \frac 12 \cos^2(2t)\ ;
\end{equation}
in other words, $\nu= \cos(2t)$. Since $d \theta_0$ is proportional to $dt$, it is also proportional to $E_1 + \bar E_1$
(see (\ref{eq:holviel})); to compute $\del \theta_0$, it is then enough to keep the part in $E_1$. We get
\begin{equation}
    v_0 = \frac12 \sin(2t) E_1 \ .
\end{equation}
It is then easy to find $\omega$:
\begin{equation}
    \omega = -\frac i{8 \psi_0} \cos(2t)\sin^2(2t)
    (\cos^2(t) (J_1 - \omega_+) - \sin^2(t) (J_2- \omega_-) )\ ;
\end{equation}
finally, the fluxes read
\begin{align}
    B&= \frac m8 \cos(2t) \left[ \sin(2t) \cos(2t) dt \wedge Da
    - \cos^2(t) J_1 + \sin^2(t) J_2
    \right]\ , \\
    F_4&= m \left[\frac12 J_{\rm FS}^2
    -\frac 1{16}\sin^3 (2t) dt \wedge Da \wedge (\cos^2(t) J_1 + \sin^2(t) J_2) \right]\ ,
\end{align}
as well as $F_0=m$.



\subsection{More general family (including ${\cal N}=3$)} 
\label{sub:inter}

We will now give a family of perturbative solutions, dual to the line of conformal field theories reviewed in section \ref{sec:ft}.

This time, it will be convenient to start, in $\cc^4$, with three holomorphic symplectic forms $\kappa_i= \kappa_{i\,AB}dz^A \wedge dz^B$, $i=1,2,3$, such that\footnote{If one identified the homogeneous coordinates with the fields in section \ref{sec:ft} as $z^A=(A_1,A_2,\bar B_1, \bar B_2)$, (\ref{eq:nui}),(\ref{eq:Winter}) would have to be written using $\kappa_i = 1_2\otimes \sigma_i$, which are hermitian but not all antisymmetric. We have preferred changing coordinates so as to use the $\kappa_i$ in (\ref{eq:omi}), which are all antisymmetric and can be identified with the coefficients of three holomorphic symplectic forms. \label{foot:chb}}
\begin{equation}\label{eq:omi}
    \kappa^A_{i\,B} \kappa^B_{j\,C}= - \delta^A{}_C \delta_{ij}
    -\epsilon_{ijk}\kappa^A_{k\,C}   \ ;
\end{equation}
namely, a holomorphic analogue of an Sp(2) structure.
One can use for example the 't Hooft symbols $\kappa_i^{AB}=
\epsilon_i{}^{AB}{}_0 +\frac12 \epsilon^{ijk}\epsilon_{ijAB}$.
From each of the $\kappa_i$ we can extract a one--form $s_{\kappa_i}\equiv s_i$ and a two--form $t_{\kappa_i}\equiv t_i$, using (\ref{eq:st}). We also introduce
\begin{equation}\label{eq:nui}
    \nu_i \equiv -\frac i {r^2}\bar z_A \kappa^A_{i\,B} z^B\ .
\end{equation}
The solution in section \ref{sub:n2} will be a particular case of the family of solutions we will present shortly, with $\kappa = \kappa_1$,
$\tilde\kappa = \kappa_2$, and $\nu = \nu_3$.

Now, when $c_1 + c_2 \neq 0$, the ${\cal N}=1$ superpotential reads
\begin{equation}\label{eq:Winter}
    W_{{\cal N}=1}= \frac12 \left(\left(\frac{2\pi}{k_1}+ \frac{2\pi}{k_2}\right)
    \nu_3^2 + (c_1 + c_2) (\nu_2^2 - \nu_1^2) \right)\ .
\end{equation}
Using the same value $T=-\frac{2 \pi}{k_1 k_2 }$ as in section \ref{sub:n2},
\begin{equation}\label{eq:thetac}
    \theta_0 = \frac12 (\nu_3^2 + c ( \nu_2^2 - \nu_1^2))
    \equiv
        \frac12 w_{ij} \nu_i \nu_j \ ,
\end{equation}
 with
\begin{equation}
     c= \frac{k_1 k_2 (c_1 + c_2)}{2\pi (k_1+k_2)}\ .
\end{equation}

We again apply the perturbative procedure we outlined in section \ref{sec:pert}.
It is no longer true (as it was in section \ref{sub:n2}) that we can use $\del_{\rm FS}\theta_0= \del \theta_0$:
\begin{equation}
    \begin{split}
        v_0 &= -\frac 12 \del_{\rm FS} \theta_0 +
        i c \,\frac{\nu_1^2}{r^2}\,{\rm Re}(\delta^{AB}\bar z_A \bar z_B \, s_1)\\
        &=\frac i2 w_{ij} \nu_i \bar z_A \kappa^A_{j\,B} Dz^B  +
        i c \,\frac{\nu_1^2}{r^2}\,{\rm Re}(\delta^{AB}\bar z_A \bar z_B \, s_1)
    \end{split}
\end{equation}
To proceed, we have to choose an SU(3) structure deformed by this $v_0$. In what follows we choose the SU(3) structure associated to the holomorphic symplectic structure $\kappa_1$: namely, $
(J_1,\Omega_1)\equiv (J_{\kappa_1}, \Omega_{\kappa_1})$.
We again compute, using (\ref{eq:oj}):
\begin{equation}
    \omega=\frac1 {2\psi_0}\left[  \bar s_1 \wedge
    (i \nu_3 s_2 +2 \theta_0\, s_1 -i c \nu_2 s_3)
    +ic\,\frac {\nu_1}{r^2}\,(\delta_{AB}z^A Dz^B \wedge \bar s_1
    + \delta^{AB} z_A z_B t_1)
     \right]\ ,
\end{equation}
and, using (\ref{eq:Bm}),
\begin{equation}\label{eq:Bn2int}
    B= m \left[
    \theta_0 J_{\rm FS}
    +\frac12 {\rm Re} (\bar s_1 \wedge (\nu_3 s_2 - c \nu_2 s_3 ) )
    +\frac{c \nu_1}{2 r^2}{\rm Re}
    (\delta_{AB}z^A (Dz^B \wedge \bar s_1 +z^B \bar t_1 ) )
    \right]\ .
\end{equation}
Following similar steps as in section \ref{sub:n2}, we get
\begin{equation}
    F_0 = m\
\end{equation}
and
\begin{equation}\label{eq:F4n2int}
    \begin{split}
        * F_4= m \Big[ &
        \frac i2 w_{ij}\bar\del_{\rm FS} \nu_i \wedge \del_{\rm FS}\nu_j
        +\frac 1{2 r^2}w_{ij}\nu_i D\bar z_A \kappa^A_{j\,B} Dz^B
        -4 \theta_0 \,J_{\rm FS}    \\
    & -\frac c {r^2}{\rm Re} ( \delta_{AB} z^A
    (- \nu_1 Dz^B \wedge \bar s_1 - \nu_1 z^B \bar t_1+ d \nu_1 \wedge \bar s_1) )
        \Big]   \ .
    \end{split}
\end{equation}
These data satisfy $d F_4 = H\wedge F_2 $, and hence define an ${\cal N}=1$ solution. As in section \ref{sub:n2}, the solution has actually ${\cal N}=2$, because one obtains the same fluxes above if one starts from the SU(3) structure $(J_2,\Omega_2)\equiv (J_{\kappa_2}, \Omega_{\kappa_2})$. This is not manifest as it was in section \ref{sub:n2}, but still true. Notice that, for generic $c$, $B$ is not invariant under exchange of $\kappa_1$ and $\kappa_2$: only $H=dB$ is.

Finally, for $c=1$ becomes invariant under an enhanced R--symmetry SO(3)$_{\rm R}$ that rotates $\kappa_1$, $\kappa_2$ and $\kappa_3$. This also implies that the solution becomes ${\cal N}=3$. This solution corresponds to the the ${\cal N}=3$ field theory we saw in section \ref{sec:ft}, for the values (\ref{eq:cn3}).

In this section, we checked the existence of a one--parameter family of infinitesimal perturbations to the ${\cal N}=6$ solution. It should be noted, however, that the existence of a supersymmetric family of solutions was guaranteed by the existence of one of them, for the following reason. The problem at first order is linear, and the difference of two solutions is a deformation that does not change $F_0$. Such a deformation is simply dual to a marginal operator in the ${\cal N}=6$ theory, and the ${\cal N}=2$ superpotential of (\ref{eq:w}) is indeed a protected operator of dimension 4 in that theory. In fact, by this argument, one can even find more general solutions by adding other marginal operators of the ${\cal N}=6$ theory to $\theta_0$ in (\ref{eq:thetac}) (for example, any $\delta w_{ij} \nu_i \nu_j$, with $\delta w_{ii}=0$). These solutions are most probably going to disappear at higher orders of perturbation theory, dual to the fact that they are not marginal operators in the family of ${\cal N}=2$ theories of section \ref{sec:ft}. Such ``spurious'' solutions will have ${\cal N}=1$ supersymmetry, unlike the ones we presented in this section, which have ${\cal N}=2$ generically and ${\cal N}=3$ for $c=1$. This extended supersymmetry appears in a non--trivial way, and is a check of the field theory predictions. 



\bigskip

{\bf Acknowledgments.} We would like to thank S.~Giombi, D.~Martelli, L.~Martucci, J.~Sparks, A.~Zaffaroni for interesting discussions.
D.~G.~is supported in part by the DOE
grant DE-FG02-90ER40542 and in part by the Roger Dashen membership
in the Institute for Advanced Study.
A.~T.~is supported in part by DOE grant DE-FG02-91ER4064.


\providecommand{\href}[2]{#2}

\end{document}